\documentstyle[aps,prd,epsf]{revtex}
\def\lsim{\raise0.3ex\hbox{$<$\kern-0.75em\raise-1.1ex\hbox{$\sim$}}}
\def\gsim{\raise0.3ex\hbox{$>$\kern-0.75em\raise-1.1ex\hbox{$\sim$}}}
\newcommand{\be} {\begin{equation}}
\newcommand{\ee} {\end{equation}}
\newcommand{\bea} {\begin{eqnarray}}
\newcommand{\eea} {\end{eqnarray}}
\def\betaquench{2.2863(10)}
\def\betacusp{1.538(46)}
\def\tcfinalfinal{171(4)}

\begin{document}

\draft
\tightenlines

\title{
%
%
\begin{flushright}
\normalsize
UTHEP-429 \\
UTCCP-P-91 \\
August 2000\\
\end{flushright}
%
%
    Phase structure and critical temperature of two-flavor QCD 
    with a renormalization group improved gauge action and
    clover improved Wilson quark action}

\author{CP-PACS Collaboration: \\[1mm]
$^1$A.~Ali Khan, $^2$S.~Aoki, $^{1,2}$R.~Burkhalter, $^1$S.~Ejiri, 
$^3$M.~Fukugita, $^4$S.~Hashimoto, $^{1,2}$N.~Ishizuka, $^{1,2}$Y.~Iwasaki, 
$^{1,2}$K.~Kanaya, $^4$T.~Kaneko, $^5$Y.~Kuramashi%
\thanks{On leave from High Energy Accelerator Research Organization(KEK),
Tsukuba, Ibaraki 305-0801, Japan},
$^1$T.~Manke%
\thanks{Present address: Dept. of Physics, Columbia University, 
538 West 120th St., New York, NY 10027, USA.},
$^1$K.~Nagai, $^2$M.~Okamoto, 
$^4$M.~Okawa, $^{1,2}$A.~Ukawa, $^{1,2}$T.~Yoshi\'e \\[2mm]}
\address{
$^1$Center for Computational Physics, University of Tsukuba, \\
Tsukuba, Ibaraki 305-8577, Japan \\
$^2$Institute of Physics, University of Tsukuba, \\
Tsukuba, Ibaraki 305-8571, Japan \\
$^3$Institute for Cosmic Ray Research, University of Tokyo, \\
Tanashi, Tokyo 188-8502, Japan \\
$^4$High Energy Accelerator Research Organization (KEK), \\
Tsukuba, Ibaraki 305-0801, Japan \\
$^5$Department of Physics, Washington University, \\
St. Louis, Missouri 63130, USA
}

\date{\today}

\maketitle

\begin{abstract}
We study the finite-temperature phase structure and 
the transition temperature of QCD with two flavors of dynamical quarks 
on a lattice with the temporal size $N_t=4$, 
using a renormalization group improved gauge action and 
the Wilson quark action improved by the clover term.
The region of a parity-broken phase is identified, 
and the finite-temperature transition line is located
on a two-dimensional parameter space of the coupling ($\beta=6/g^2$) and
hopping parameter $K$.
Near the chiral transition point, defined as the crossing point of the 
critical line of the vanishing pion mass and the line of finite-temperature 
transition, the system exhibits behavior well described 
by the scaling exponents of the 
three-dimensional O(4) spin model.  This indicates a second-order 
chiral transition in the continuum limit. 
The transition temperature in the chiral limit is estimated to be
$T_c = \tcfinalfinal$ MeV.
\end{abstract}

\newpage

\section{Introduction}
\label{sec:intro}

With the development of recent experimental projects toward a detection 
of the high-temperature quark-gluon plasma in relativistic heavy ion 
collisions, it is urgent to theoretically establish 
thermal properties of QCD directly from its first principles.

For the pure gluon system, numerical simulations on the lattice are quite 
advanced by now, with the critical temperature and the equation of state, 
which are the basic thermodynamic quantities needed for phenomenological 
studies, well established in the continuum limit\cite{Karsch99}. 
Full lattice QCD simulations incorporating dynamical quarks, however, 
have not reached such a status.  In particular studies on spatially 
large lattices, which are required for determination of equation of state, 
are still limited to lattices of a small temporal size, $N_t=4$ and 
6\cite{MILC-EOS}, due to large computational power needed for 
dynamical quark simulations.  

A possible way to bring the simulation close to the continuum 
is to employ improved 
lattice actions: the continuum limit may be taken 
for thermodynamic quantities from coarse lattice spacings 
and correspondingly small temporal lattice sizes such as $N_t=4$ and 6 
or even smaller.
For the pure gluon system, there are a number of tests of this idea 
reported some time ago\cite{Ukawa96}. 
More recently it is shown\cite{Okamoto} that the equation of state 
calculated for a  renormalization-group (RG) improved gluon 
action\cite{Iwasaki83} agrees with that for the plaquette 
action\cite{Bielefeld} in the continuum limit. 
In the present work we pursue 
an extension of this pure gluon study to full QCD 
with two flavors of dynamical quarks\cite{prelim}.
 
Here we focus our attention to the Wilson-type quark action, and
study the effect of improvement for both gluon and quark sectors.
Compared to Kogut-Susskind (staggered) quark action\cite{Bielefeld-EOS}, 
a clear advantage of the Wilson-type quark actions is manifest 
flavor symmetry.  On the other hand, chiral symmetry 
is explicitly broken causing some subtleties in the analysis. 
This leads to large lattice artifacts at the range of lattice spacing 
$a^{-1}\approx 1$--2 GeV extensively used in finite 
temperature simulations, when the combination of 
the plaquette gluon action and the standard Wilson quark action is used.
The most notable artifact 
is an unexpected strengthening of the finite-temperature 
transition at intermediate values of 
the quark mass \cite{MILC,Tsukuba96}. 
This undesired behavior disappears when an RG-improved gauge action is 
employed \cite{Tsukuba97}.
A comparison with other action combinations, 
the plaquette gluon action and
a clover-improved \cite{SWclover} Wilson quark action \cite{cloverFT},
or the Symanzik-improved gluon action and 
a clover-improved quark action \cite{MILC97}, 
indicates that improvement of the gluon action 
is essential in removing the dominant lattice artifacts at finite 
lattice spacings. On the other hand, we expect that 
improvement for the quark action renders the behavior of
fermionic thermodynamic quantities close to continuum.
 
Previous studies with the standard Wilson action 
have shown that the phase diagram in the plane of coupling and hopping 
parameter $(\beta=6/g^2, K)$ contains a phase with spontaneous breakdown 
of parity-flavor symmetry\cite{Aoki}.  
The boundary of this phase is characterized by vanishing of pion (screening) 
mass. The interplay of 
the boundary with the line of finite-temperature transition 
plays a crucial role in understanding the finite-temperature 
transition with Wilson-type quark actions \cite{AUU96,Tsukuba96,Tsukuba97a}. 
Thus, prior to calculation of thermodynamics quantities, an examination of 
the finite-temperature phase structure is indispensable 
when one uses the clover-improved 
Wilson quark action coupled to the RG-improved gluon action.
  
A particularly important point concerns the universality property 
of the finite-temperature transition.
For the standard staggered quark action, extensive scaling 
analyses carried out for $N_t=4$ failed to establish 
either the O(4) scaling expected from the effective sigma 
model analysis\cite{PisarskiWilczek} or the O(2) scaling suggested from
the actual symmetry of the staggered quark 
action\cite{KL,O4JLQCD,O4Bielefeld,O4MILC}.
On the other hand, a study with the standard Wilson quark action 
coupled to the RG-improved gauge action found that 
the O(4) scaling ansatz describes well the chiral condensate 
data \cite{Tsukuba97,Tsukuba97a}.
Our interest here is to explore whether the O(4) scaling is maintained 
for the clover-improved Wilson action. 
Such an analysis is also necessary to quantitatively locate 
the point of transition for massless quark.  Combining this information 
with measurements of hadron masses at zero temperature, 
we obtain an estimate of the critical temperature. 

In this paper we adopt specifically the improved 
gluon and quark actions which is resulted from 
a systematic comparative study of 
various combinations of improved actions\cite{comparative}.
The combination is the same 
as that we have used in our zero-temperature studies 
reported elsewhere\cite{CPPACSfull,CPPACSmq}.

This paper is organized as follows. 
The action and the simulation parameters are described in 
Sec.~\ref{sec:simulations}. 
The phase structure for our action is studied in Sec.~\ref{sec:phase}. 
Section \ref{sec:o4} is devoted to the issue of the O(4) universality.
Assuming the O(4) critical exponents, we then extrapolate the transition 
line to obtain a precise estimate of the chiral transition point. 
The transition temperature in physical units is discussed in 
Sec.~\ref{sec:scalingT}.
A brief conclusion is given in Sec.~\ref{sec:conclusion}.

\section{Simulation}
\label{sec:simulations}

\subsection{Choice of the action}

We employ the RG-improved gluon action \cite{Iwasaki83} 
\be
S_g = - \beta  \{\;c_0\sum_{x,\mu <\nu} W_{\mu\nu}^{1\times 1}(x) 
+c_1\sum_{x,\mu ,\nu} W_{\mu\nu}^{1\times 2}(x)\},
\label{eq:6LinkAct} 
\ee
with $\beta=6/g^2$ and $c_1 = -0.331$ and $c_0 = 1 - 8c_1$, 
coupled with the clover-improved Wilson quark action \cite{SWclover}
defined by 
\begin{eqnarray}
S_q & = & \sum_{x,y}\overline{q}_x D_{x,y}q_y,\\
D_{x,y} & = & \delta_{xy}
- K \sum_\mu \{(1-\gamma_\mu)U_{x,\mu} \delta_{x+\hat\mu,y}
      + (1+\gamma_\mu)U_{x,\mu}^{\dag} \delta_{x,y+\hat\mu} \}
- \, \delta_{xy} c_{SW} K \sum_{\mu < \nu}
         \sigma_{\mu\nu} F_{\mu\nu} .
\end{eqnarray}
Here $F_{\mu\nu}$ is the lattice discretization of the field strength,
\be
F_{\mu\nu} = \frac{1}{8i} (f_{\mu\nu} - f_{\mu\nu}^{\dag}),
\label{eq:CloverLeaf}
\ee
with $f_{\mu\nu}$ the standard clover-shaped combination of gauge links, 
and we adopt the meanfield-improved clover coefficient 
\be
c_{SW} = (W^{1\times 1})^{-3/4} = (1-0.8412 \beta^{-1})^{-3/4},
\label{eq:csw} 
\ee
using $W^{1\times 1}$ calculated in the one-loop perturbation 
theory\cite{Iwasaki83}.
This choice of $c_{SW}$, when expanded in $\beta$, agrees well with 
the actual one-loop result 
$
c_{SW} = 1 + 0.678(18)/\beta + \cdots
$
\cite{csw-one-loop}.
Furthermore, the one-loop values of $W^{1\times 1}$ reproduce the 
mean values of plaquette within 8\% for the range of $\beta$ and $K$ 
of our studies\cite{CPPACSfull,CPPACSmq}. 

This combination of improved gauge and quark actions was tested 
in our full QCD comparative study\cite{comparative}.  
We found that the scaling violation in light hadron masses and 
the violation of rotational invariance in a static quark potential 
are both small with this combination of actions 
already at $a^{-1}\approx 1$~GeV 
where our finite temperature simulations are made, 
as compared to $a^{-1} \gsim\, 2$~GeV needed  
for the standard plaquette gauge and Wilson quark actions.

\subsection{Details of simulations}

We make our finite-temperature simulations on a lattice with a 
temporal extension $N_t=4$.  Zero-temperature simulations with 
temporal sizes comparable or larger than spatial sizes 
are also made to calculate hadron masses 
for the purpose of fixing the scale.

Our simulation parameters are summarized in 
Tables~\ref{tab:param16x4}--\ref{tab:param12x24}.
Runs are carried out on lattices 
of size $16^3\times 4$ and $16^4$ at $\beta = 1.8$--2.2, 
and of size $8^3\times 4$ and $12^3 \times 24$ at $\beta = 1.2$--1.7. 
Except for the case of the $12^3 \times 24$ lattice, 
we use anti-periodic boundary conditions in the temporal direction 
for quarks. All other boundary conditions are taken to be periodic.
Simulations are carried out at seven to eleven values of the hopping
parameter $K$ for each $\beta$, which correspond to 
$m_{\rm PS}/m_{\rm V}\approx 0.4$--0.9 for the $8^3\times 4$ lattices, 
and $0.6$--1.0 for the $16^3\times 4$ lattices. 

The Hybrid Monte Carlo algorithm is employed to generate 
full QCD configurations with two flavors of dynamical quarks. 
Details of the simulation are basically the same as in our zero temperature 
studies \cite{CPPACSfull,CPPACSmq}. 
The molecular dynamics time step $\delta\tau$ is chosen to yield
an acceptance rate greater than about 80\%. 
The length of one trajectory is unity in most cases, but is reduced
down to 0.25 at several points close to the critical line to keep
the acceptance rate of about 80\% within 200 molecular dynamics time steps.
The inversion of the quark matrix is made with the 
BiConjugate Gradient Stabilized (BiCGStab) method 
for $K < K_c$, 
where $K_c$ is the critical hopping parameter at zero temperature where 
pion mass vanishes.
For $K > K_c$, we use the Conjugate Gradient (CG) method 
since the clover-Wilson operator 
may have negative eigenvalues, in which case the BiCGStab algorithm 
fails to converge.  

We measure gluonic observables, Wilson loops and the Polyakov 
line, at every trajectory. 
Hadron (screening) propagators are calculated at every 2--5 trajectories 
using both a point source and an exponentially smeared source for quark, 
and a point sink. 
For hadronic measurements on the $8^3\times4$ lattice, we double 
periodically the lattice in one of the spatial directions.
On the $8^3\times 4$ and $12^3\times24$ lattices, we determine hadron 
masses by a single hyperbolic cosine fit to the propagator calculated 
with the smeared source, because of its clear plateaus of effective mass. 
On the $16^3\times 4$ and $16^4$ lattices, since a plateau is sometimes 
less clear 
with our smeared source, 
we perform a combined fit using both point and smeared sources. 

We measure the current quark mass defined through an axial vector
Ward-Takahashi identity \cite{Itoh,Bochicchio},
\be
\nabla_\mu A_\mu = 2m_q P + O(a)
\ee
where $P$ is the pseudoscalar density and $A_\mu$ the $\mu$-th
component of the local axial vector current.  
In practice we make a simultaneous fit of the two-point functions 
$\langle A_\mu(t)P(0) \rangle$ and $\langle P(t)P(0) \rangle$ 
to extract the pion mass $m_\pi$ 
and the amplitudes $\langle 0|P|\pi(\vec{p}=0)\rangle$, 
$\langle 0|A_\mu|\pi(\vec{p}=0)\rangle$, from which we compute
\be
m_q =-m_\pi \frac{\langle\,0\,|\,A_\mu\,|\,\pi(\vec{p}=0)\,\rangle}
	         {\langle\,0\,|\,P\,|\,\pi(\vec{p}=0)\,\rangle}.
\label{eq:mq}
\ee
We choose $\mu=4$ for 
zero temperature simulations, and $\mu=3$ at finite temperatures 
for which the screening masses are determined along the $z$ axis.
We also use an alternative definition of quark mass given by 
\be
m_q = (1/2a)\,(K^{-1} - K_c^{-1}).
\ee
While these different definitions of $m_q$ give different values
at finite $\beta$, they converge to the same value in the continuum 
limit\cite{CPPACSq,CPPACSmq}.
In this work, we concentrate on the phase structure and critical 
properties around the finite temperature chiral transition point. 
Therefore, in this work, we ignore the renormalization factors $Z_P$, 
$Z_A$ and $Z_m$ since they are regular around the transition point. 

We use the zero-temperature vector meson mass $m_{\rm V}$ as a measure of the 
lattice scale.
For $\beta\geq 1.8$, detailed hadron mass data for light quarks are 
available from our extensive calculations\cite{CPPACSfull}.  
We use the results of hadron masses obtained in this reference for our 
analysis.
These data, however, do not fully cover the region of heavy quark 
where the finite-temperature transition is located for $N_t=4$ at 
$\beta\geq 1.8$.  
We evaluate hadron masses in these regions, 
and interpolate them by a cubic spline formula to the point of transition. 

\begin{figure}[tb]
  \begin{center}
    \leavevmode
    \epsfxsize=11cm 
    \epsfbox{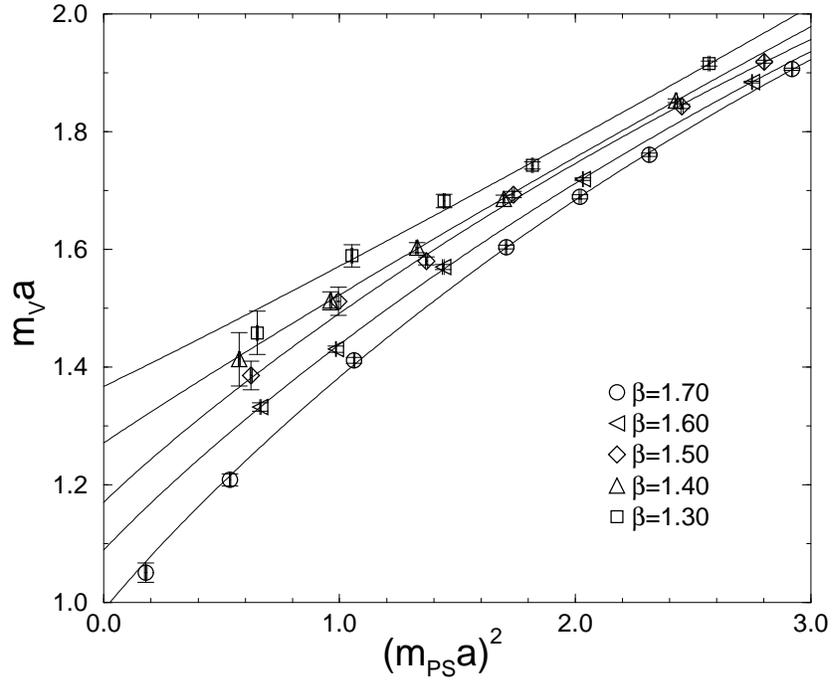}
    \caption{Chiral extrapolation of vector meson mass as a function 
    of $(m_{\rm PS}a)^2$ obtained on $12^3\times 24$ lattice.}
    \label{fig:rhoM_12x24}
  \end{center}
\end{figure}

No previous data are available in the region of strong coupling 
$\beta=1.3$--1.7. Thus the hadron masses in this region
are measured in the present work. 
We fit the vector meson mass in terms of pseudoscalar meson mass 
$m_{\rm PS}$ by an ansatz inspired by chiral perturbation theory,  
\begin{equation}
m_{\rm V}a=A_{\rm V}+B_{\rm V}(m_{\rm PS}a)^2+C_{\rm V}(m_{\rm PS}a)^3.
\label{eq:mrho}
\end{equation}
The fit curves are illustrated in Fig.~\ref{fig:rhoM_12x24}. 
The lattice spacing defined by identifying 
$m_{\rm V}(T=0) = m_\rho = 770$ MeV 
in the limit of zero pion mass is given in Table~\ref{tab:Kc}.

Errors are determined by a jack-knife method. 
From a study of bin-size dependence, we adopt the bin
size of 1--10 configurations ($\it i.e.$ 2--50 trajectories) for
hadron masses and 5--20 trajectories for other gluonic quantities.

\section{Phase structure}
\label{sec:phase}

The phase diagram obtained in our study is summarized in Fig.~\ref{fig:phase}. 
Details are discussed in the following subsections.
Thin open symbols show the simulation points on the $N_t=4$ lattices.
The solid line connecting symbols 
denoted as ``$K_c(T=0)$'' represents the critical line 
of vanishing pion mass at zero temperature.
Other objects on the phase diagram are obtained on the $N_t=4$ lattices.
The $K_t$-line shows the location of the finite-temperature 
transition, above which the system is in the high-temperature phase.
We expect the $K_t$-line to cross the $K_c(T=0)$ line.  The crossing 
point is a natural candidate to be identified as the 
point of chiral transition\cite{Tsukuba96}.  
The shaded region shows the parity-broken phase \cite{Aoki}.
The lower boundary of the parity-broken phase lies closely 
above the $K_c(T=0)$ line.

\subsection{Critical line at zero temperature}
\label{sec:chirallimit}

\begin{figure}[tb]
  \begin{center}
    \leavevmode
    \epsfxsize=11cm
    \epsfbox{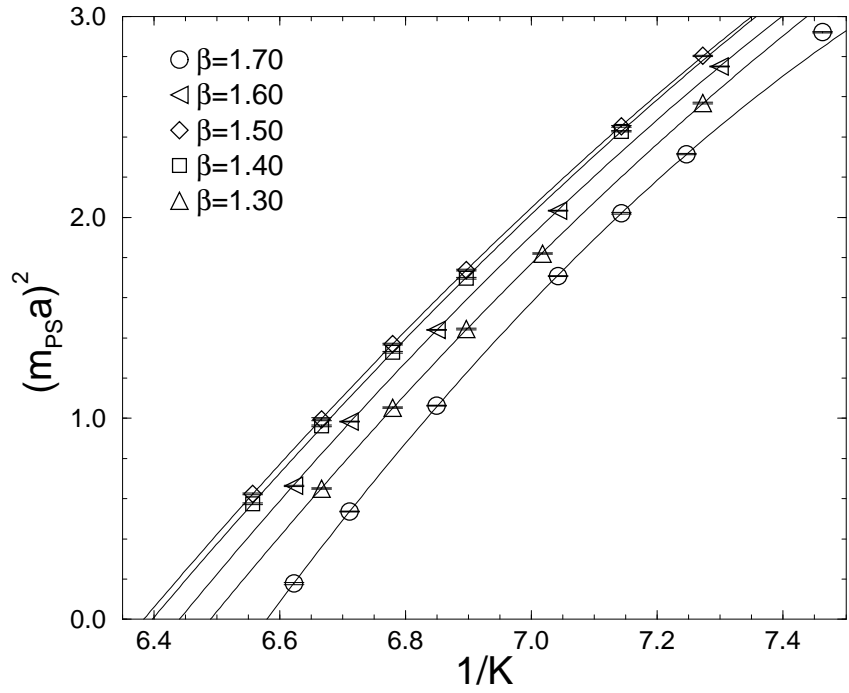}
    \caption{Chiral extrapolation of pseudo scalar meson mass as a function
    of $K$ on $12^3\times 24$ lattice.}
    \label{fig:piM2_12x24}
  \end{center}
\end{figure}

The location of the critical line at zero temperature  
$K_c(T=0)$ where 
the pseudoscalar meson mass vanishes is essential information 
for discussion of chiral properties with Wilson-type quark actions. 
To determine $K_c(T=0)$, we extrapolate $m_{\rm PS}^2$ 
as a function of $1/K$ using a quadratic ansatz:
\be
<(m_{\rm PS}a)^2 =   B_{\rm PS}\left(\frac{1}{K}-\frac{1}{K_c}\right)
                  + C_{\rm PS}\left(\frac{1}{K}-\frac{1}{K_c}\right)^2.
\label{eq:mpi}
\ee
The results for $m_{\rm PS}^2$ are illustrated  
in Fig.~\ref{fig:piM2_12x24} in the strong-coupling region of $\beta=1.3-1.7$.
Our values for $K_c(T=0)$ are summarized in Table~\ref{tab:Kc} and 
plotted in Fig.~\ref{fig:phase}.

Towards the weak coupling limit, $\beta=\infty$, the $K_c(T=0)$-line 
gradually approaches the free Wilson quark value 1/8. 
For the case of the standard unimproved Wilson quark action, 
$K_c(T=0)$ is a monotonically decreasing curve connecting
$K\approx 1/4$ at $\beta=0$ and 1/8 at $\beta=\infty$.
For the case of our clover-improved quark action, 
the $K_c(T=0)$-line shows a maximum in $K$ at $\beta \approx 1.47$,  
and decreases as we lower $\beta$ below this value. 
This is likely to arise from our choice (\ref{eq:csw}) 
for the clover coefficient,  
which diverges as $\beta$ approaches 0.8412.

\subsection{Finite-temperature transition}
\label{sec:FTtransition}

\begin{figure}[tb]
  \begin{center}
    \leavevmode
    \epsfxsize=11cm 
    \epsfbox{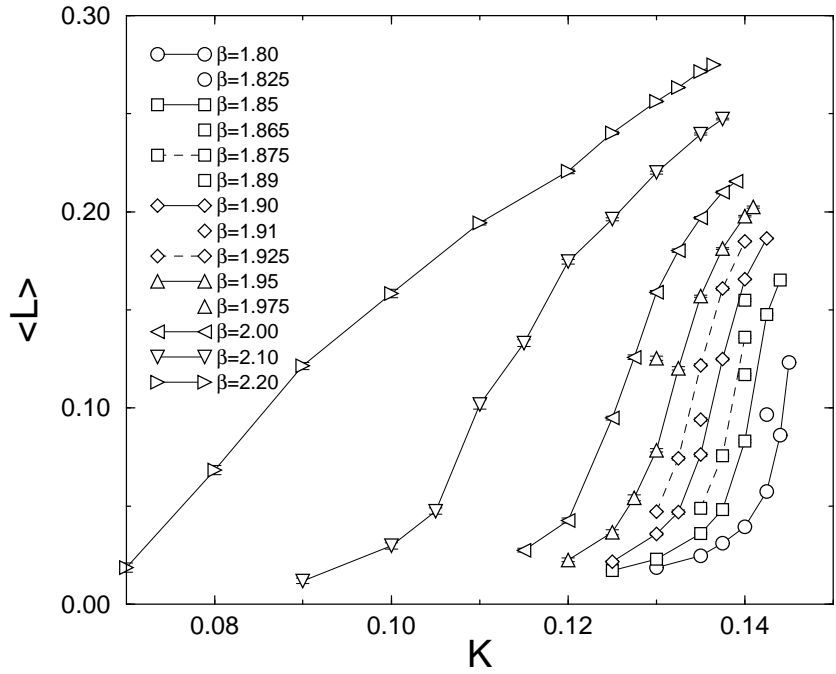}
    \caption{Polyakov line obtained on $16^3\times 4$ lattice.}
    \label{fig:pline16x4}
  \end{center}
\end{figure}

\begin{figure}[tb]
  \begin{center}
    \leavevmode
    \epsfxsize=11cm 
    \epsfbox{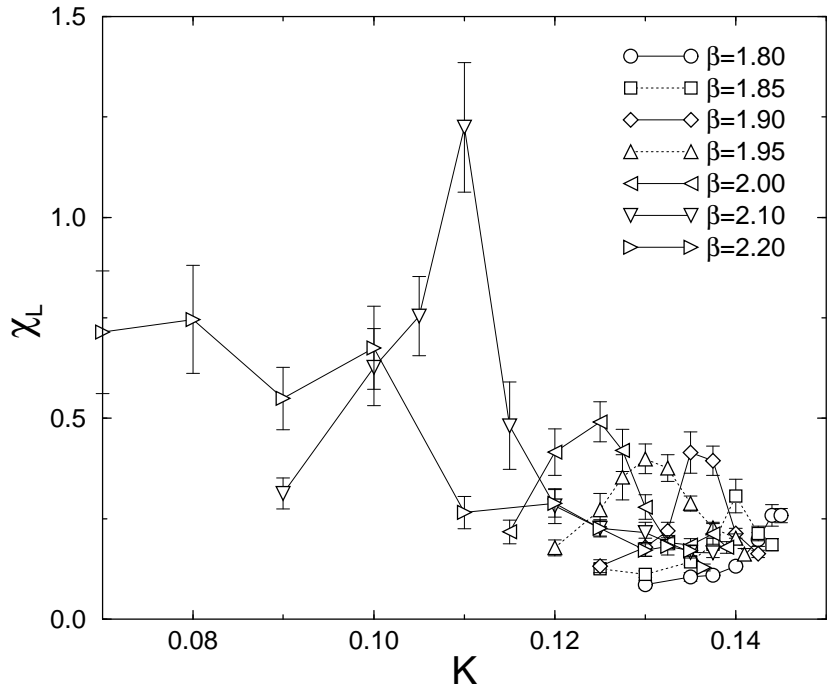}
    \caption{Polyakov line susceptibility obtained on 
      $16^3\times 4$ lattice. Several data are omitted for clarity 
      of the plot.}
    \label{fig:psus16x4}
  \end{center}
\end{figure}

We identify the finite temperature transition point
$K_t$ from inspection of the Polyakov line and Wilson loops. 
In Figs.~\ref{fig:pline16x4} and \ref{fig:psus16x4} 
the Polyakov line and its susceptibility at $\beta=1.8$--2.2
on the $16^3\times 4$ lattice are shown.   
We fit the peak of the susceptibility by a Gaussian form using 
3 or 4 points near the peak, except for $\beta=2.2$.  
The results for $K_t$ from the fit 
are given in Table~\ref{tab:Kt}.  
The ratios $m_{\rm PS}/m_{\rm V}$ and $T_{pc}/m_{\rm V}$ 
with zero-temperature meson masses interpolated to the $K_t$ point
are also summarized in the table, where $T_{pc}$ is the 
pseudocritical temperature at $K_t$.
These results are used in Secs.~\ref{sec:scalingB}
and \ref{sec:scalingT}.

On the $K_t$-line thus determined, other physical observables, 
such as the plaquette, also show rapid changes. 
Smoothness of the data around $K_t$ suggests, however, that,
for the range of quark mass we studied, 
the finite-temperature transition at $K_t$ is an analytic crossover.
From Fig.~\ref{fig:pline16x4} and similar plots for
other observables, we see that 
the crossover becomes monotonically weaker with increasing $\beta$.
Increasing $\beta$ on the $K_t$ line corresponds to increasing the 
distance to the critical line, i.e. increasing the quark mass.
In the limit of infinite quark mass $K=0$, the $K_t$ line will end at  
the first-order deconfinement transition of pure gauge theory,
which is located at $\beta=\betaquench$ for $N_t=4$ 
\cite{kanekoTc,Okamoto}. 

\begin{figure}[tb]
  \begin{center}
    \leavevmode
    \epsfxsize=11cm
    \epsfbox{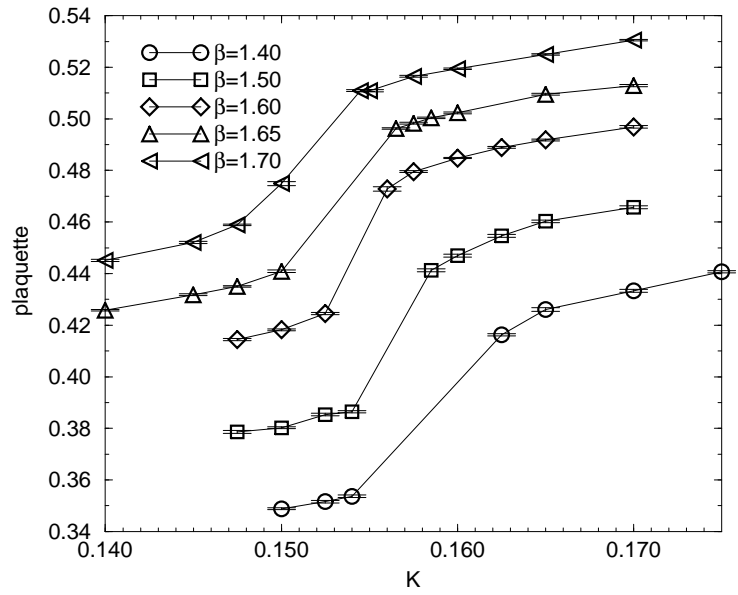}
    \caption{Spatial plaquette obtained on $8^3\times 4$ lattice.}
    \label{fig:plaq8x4}
  \end{center}
\end{figure}

For the smaller $\beta$ region ($\beta \leq 1.7$) where simulations are 
made on an $8^3\times 4$ lattice, data for the 
Polyakov line become too noisy to determine the position of $K_t$.
We alternatively take the spatial plaquette shown in Fig.~\ref{fig:plaq8x4}, 
and identify $K_t$ from its rapid changes. 
The range of $K_t$ for $\beta=1.6$--1.7 thus estimated 
is summarized in Table~\ref{tab:Kt}.

\begin{figure}[t]
  \begin{center}
    \leavevmode
    \epsfxsize=14cm 
    \epsfbox{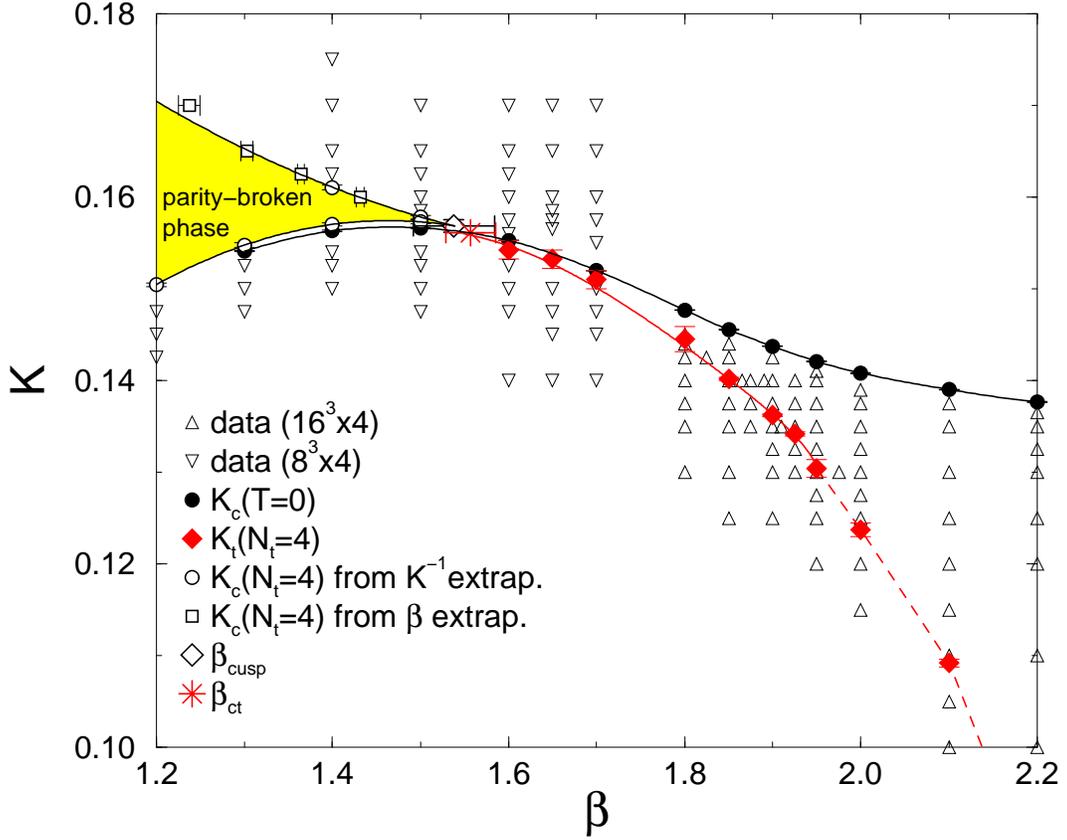}
    \caption{Phase diagram for the RG-improved gauge action and
    clover quark action at $N_t=4$.}
    \label{fig:phase}
  \end{center}
\end{figure}

The $K_t$-line approaches the zero-temperature critical line $K_c(T=0)$
as $\beta$ decreases (see Fig.~\ref{fig:phase}).   
We expect the two lines to cross at a point $(\beta_{ct}, K_{ct})$. 
This point is a natural candidate for the point of chiral transition 
since the pion mass, and hence also the quark mass defined through the 
Ward-Takahashi identity, vanishes at zero temperature\cite{Tsukuba96}.  
The location of this point obtained by scaling analyses 
(see Sec.~\ref{sec:scalingB}) is marked by a star in Fig.~\ref{fig:phase}. 

Our estimate for $K_t$ stops at $\beta=1.6$.  For $\beta \lsim 1.5$, 
because the parity-broken phase appears at 
intermediate $K$ as discussed in the next subsection, a more 
detailed study is required to identify the location of $K_t$. 

\subsection{Parity-broken phase}
\label{sec:PBphase}

For Wilson-type quark actions, chiral symmetry is explicitly 
broken by the Wilson term away from the continuum limit. 
The appearance of massless pseudoscalar meson (pion) at $K_c(T=0)$ is, 
therefore, not due to spontaneous breakdown of chiral symmetry. 
Analytic calculations in the strong coupling limit 
and numerical simulations at intermediate couplings show 
that there exists a region in the phase diagram in which 
parity-flavor symmetry is spontaneously broken 
(``parity-broken phase'') \cite{Aoki,AUU96}.
In this picture, the pion is understood as the zero mode of 
a second-order phase transition that takes place along the $K_c(T=0)$ line, 
signaling spontaneous breakdown of parity-flavor symmetry. 

At zero temperature, the parity-broken phase extends from the 
strong coupling region towards the weak coupling region, 
forming sharp cusps which touch the weak coupling limit $\beta=\infty$ at 
$K=1/8$ and at four other values of $K$. 
In the present work, we concentrate on the branch of the parity-broken 
phase with the smallest positive values of $K$. 
The lower boundary of this branch is 
the usual critical line denoted by $K_c(T=0)$ in Fig.~\ref{fig:phase}.

For finite temporal size $N_t$, the cusp of the parity-broken phase retracts 
from the weak coupling limit to a finite value of $\beta$. 
Let us denote the position of the cusp as 
$(\beta_{\rm cusp}, K_{\rm cusp})$,  
and the boundary of the parity-broken phase as $K_c(N_t)$.  
This boundary is double-valued for $\beta\leq\beta_{cusp}$, whose two 
branches we denote as $K_c^{\rm lower}(N_t)<K_c^{\rm upper}(N_t)$.   

In the low temperature phase $K<K_t$, we expect a massless pion to appear as 
$K$ is increased.
Therefore, the lower part of the $K_c(N_t)$-line should be located 
near the zero-temperature critical line $K_c(T=0)$, 
with the difference being $O(a)$.
On the other hand, we do not expect a massless pion in the high 
temperature phase $K>K_t$.  
Hence the whole of $K_c(N_t)$-line should be in the low 
temperature phase.
These considerations imply that the $K_t$-line runs closely past the cusp 
of the $K_c(N_t)$ line.  Therefore we expect that the  
the chiral transition point is located near and above the cusp\cite{AUU96}:
\be
\beta_{ct} = \beta_{\rm cusp} + O(a),
\label{eq:ctgecusp}
\ee
with $\beta_{ct} \ge \beta_{\rm cusp}$.  
The possibility that the chiral transition point actually agrees with 
the cusp is not excluded. 
For the case of the standard Wilson quark action coupled with 
the plaquette gauge action, 
$\beta_{ct}$ has indeed been 
found to be close to $\beta_{\rm cusp}$ \cite{AUU96}. 
Similar results were obtained for the case of 
the RG-improved gauge action with the standard Wilson quark action 
\cite{Tsukuba97a}. 

\begin{figure}[tb]
  \begin{center}
    \leavevmode
    \epsfxsize=11cm 
    \epsfbox{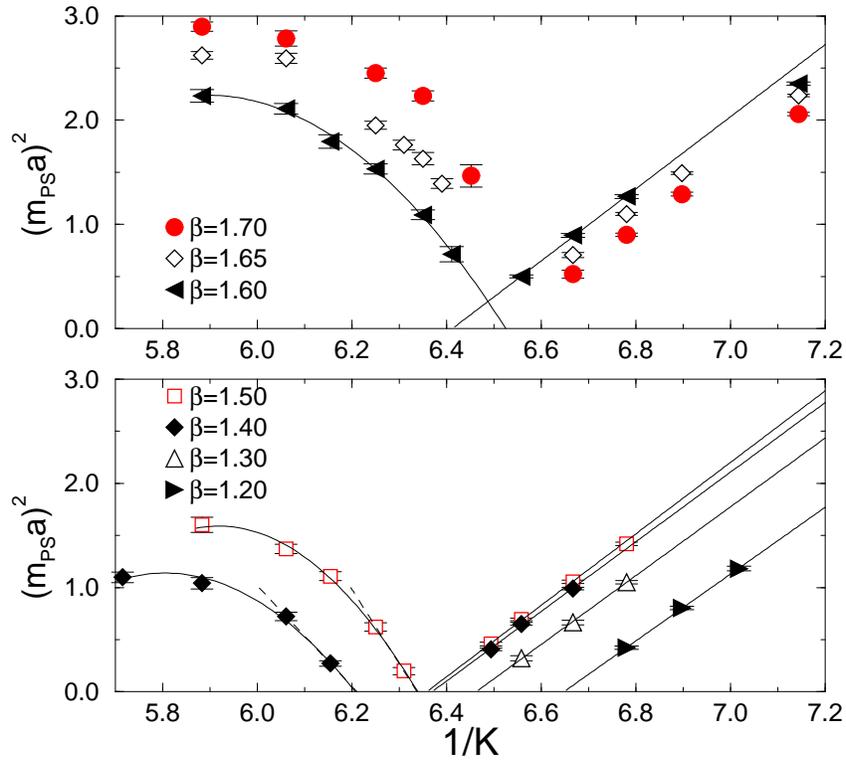}
    \caption{Pseudoscalar meson mass as a function 
    of $1/K$ obtained on $8^3\times 4$ lattice.}
    \label{fig:piM2_8x4}
  \end{center}
\end{figure}

Figure~\ref{fig:piM2_8x4} shows results for pion screening mass 
squared $(m_{\rm PS}a)^2$ with our action on an $N_t=4$ lattice.
Solid lines are linear or quadratic fits in $1/K$. 
In the upper figure showing results for $1.7\geq \beta\geq 1.6$, 
the two fit lines for $\beta=1.6$, from large or small values of 
$1/K$, cross each other.  This indicates that $m_{\rm PS}$ remains 
finite for all values of $K$, and hence 
the parity-broken phase does not exist for $\beta \ge 1.6$.

In contrast, when we decrease $\beta$ down to 1.5 and 1.4, 
we see evidence for two values of $K_c(N_t=4)$ (lower figure in 
Fig.~\ref{fig:piM2_8x4}) corresponding to the upper and lower 
boundaries of the cusp of the parity-broken phase. 

\begin{figure}[tb]
  \begin{center}
    \leavevmode
    \epsfxsize=11cm 
    \epsfbox{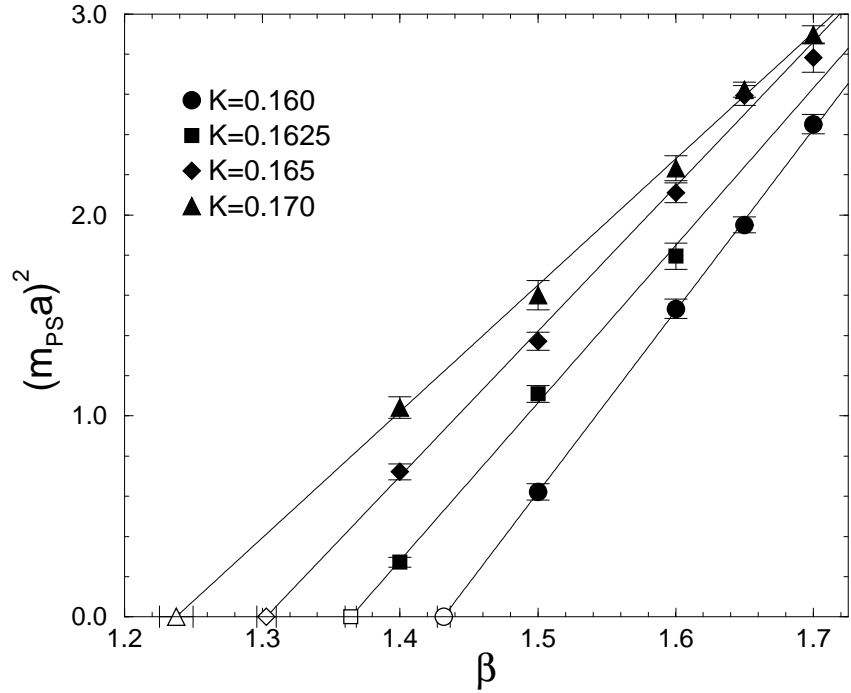}
    \caption{Pseudo scalar meson mass as a function 
    of $\beta$ obtained on $8^3\times 4$ lattice.}
    \label{fig:extrpKc2}
  \end{center}
\end{figure}

We determine the location of $K_c(N_t=4)$ by an extrapolation of 
$(m_{\rm PS} a)^2$ linearly in $1/K$, using the lightest two to three
points.
We confirm that a quadratic extrapolation gives consistent results.  
Results of $K_c(N_t=4)$ are summarized in Table~\ref{tab:Kc}.
We find the gap between two values of $1/K_c(N_t=4)$ to be
\begin{eqnarray}
\Delta (1/K_c) &=& 0.020(12) {\rm \;\;\ at \ } \beta = 1.5\\
               &=& 0.159(12) {\rm \;\;\ at \ } \beta = 1.4.
\end{eqnarray}
For the upper $K_c(N_t=4)$ line, 
we also try to estimate its location by extrapolating  
$(m_{\rm PS} a)^2$ linearly in $\beta$, 
at fixed $K=0.16$, 0.1625, 0.165, and 0.17, 
as shown in Fig.~\ref{fig:extrpKc2}.
Results are listed in Table~\ref{tab:Kc}. 
As is seen in Fig.~\ref{fig:phase}, 
the points for $K_c(N_t=4)$,  obtained from the $1/K$ fits and the
$\beta$ fits, together form a smooth curve. 

\begin{figure}[tb]
  \begin{center}
    \leavevmode
    \epsfxsize=11cm 
    \epsfbox{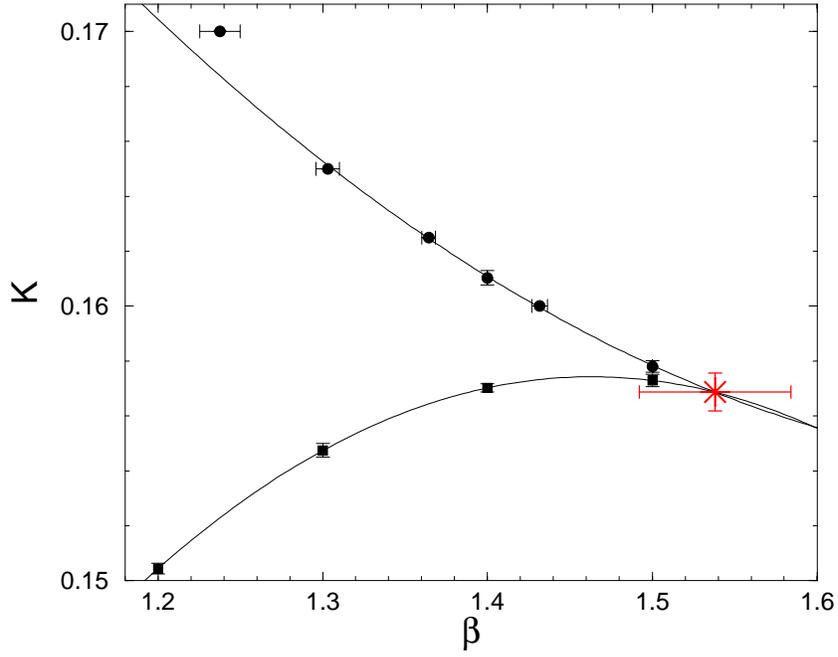}
    \caption{Cusp of the parity-broken phase. Cross shows an estimate
             for the crossing point of the two lines extrapolated 
             quadratically in $\beta$. }
    \label{fig:cusp}
  \end{center}
\end{figure}

An enlargement of the phase diagram around the parity-broken phase 
is given in Fig.~\ref{fig:cusp}. 
Since the parity-broken phase is absent at $\beta=1.6$ the boundaries of 
the parity-broken phase should terminate at a cusp
at $\beta = 1.5$--1.6.
In order to determine the location of the cusp point more precisely, 
we fit the upper and lower $K_c(N_t=4)$ data separately 
by a quadratic ansatz.
From these fits, we obtain
\be
\beta_{\rm cusp} = \betacusp, \;\;\;\;
K_{\rm cusp}    = 0.15687(69),
\label{eq:cusp}
\ee
where the errors are estimated from the crossing points of the one standard 
deviation error bands of the two fits. 
From the discussions around (\ref{eq:ctgecusp}), 
this value for $\beta_{\rm cusp}$ gives a lower bound 
of $\beta_{ct}$.

\section{O(4) scaling of chiral condensate}
\label{sec:o4} 

From universality arguments, the finite-temperature QCD transition 
near the chiral limit is expected to 
be described by an effective $\sigma$ model \cite{PisarskiWilczek}.
According to this description, the transition for two quark flavors 
in the chiral limit is either first order or second order 
depending on the strength of anomalous coupling 
which breaks axial U$\!_A$(1) symmetry. 
When this effect is negligible, the transition is of second order, and 
one expects that the critical properties are described by those of 
the three-dimensional O(4) Heisenberg model. 
Hence tests of O(4) scaling provides us with a useful way to study  
universality properties of the chiral transition for two-flavor QCD 
\cite{KL}.

In the O(4) Heisenberg model, the order parameter is given by 
the magnetization $M$. 
Near the second order transition point, $M$ satisfies the 
following scaling relation;
\begin{equation}
M / h^{1/\delta} = f(t/h^{1/\beta\delta}).
\label{eq:o4}
\end{equation}
where $h$ is the external magnetic field, 
$t=[T-T_c(h=0)]/T_c(h=0)$ is the reduced temperature, 
the exponents have the values $1/\beta\delta = 0.537(7)$ and 
$1/\delta = 0.2061(9)$ \cite{O4exponent}, and  
the O(4) scaling function $f(x)$ is also known \cite{Toussaint}.

In Ref.~\cite{Tsukuba97},
it was shown that the O(4) scaling of (\ref{eq:o4}) is well satisfied 
for the standard Wilson quark action combined with the RG-improved 
gauge action under the identification  
$t \sim \beta - \beta_{ct}$, $h \sim m_q a$, and
$M \sim \langle \bar\psi \psi\rangle$,  where quark mass defined 
by the axial vector Ward-Takahashi identity was employed. 
It is important to note that 
the naive definition of the chiral condensate 
$\langle \bar{\psi} \psi \rangle$
is not adequate for Wilson-type quarks 
because chiral symmetry is explicitly broken. 
A proper subtraction and renormalization are required. 
A properly subtracted 
$\langle \bar{\psi} \psi \rangle$ can be defined via an axial 
Ward-Takahashi identity \cite{Bochicchio} as first employed in Ref.~\cite{Tsukuba97}:
\begin{equation}
\langle \bar{\psi} \psi \rangle_{\rm sub} 
= 2 m_q a Z \sum_x \langle \pi(x) \pi(0) \rangle
\label{eq:PBPsub}
\end{equation}
where $\pi(x) = \bar{q}_x \gamma_5 q_x$ with $q_x$ the lattice quark 
field.
For the normalization coefficient $Z$ we adopt the tree value 
$Z=(2K)^2$, which is sufficient for our study of critical properties because 
$Z$ is regular at the finite temperature transition point.

\begin{figure}[tb]
  \begin{center}
    \leavevmode
    \epsfxsize=11cm 
    \epsfbox{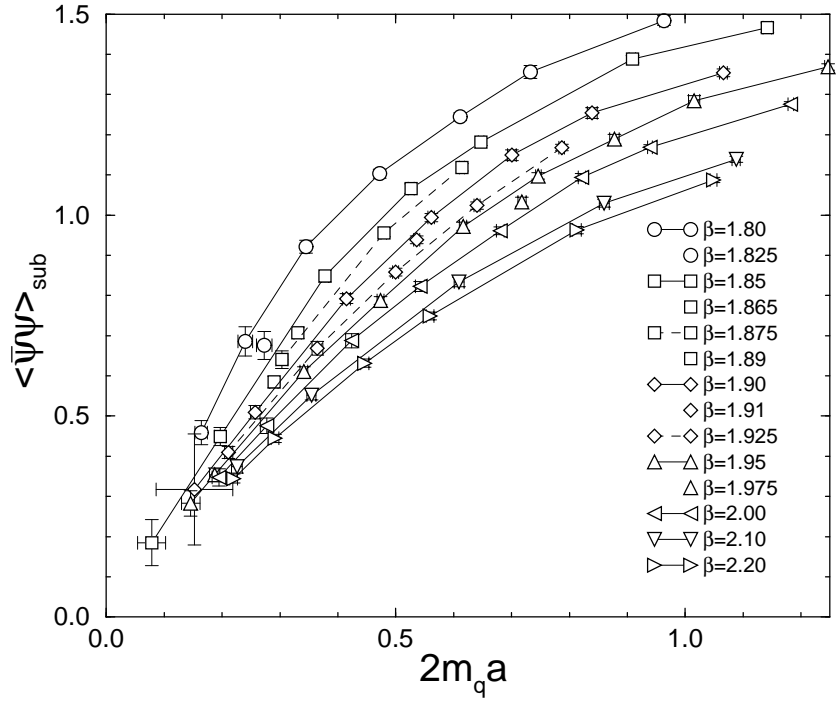}
    \caption{$\langle \bar{\psi}\psi\rangle_{sub}$ as a
    function of $2m_qa$ on $16^3\times 4$ lattice.}
    \label{fig:pbp-2mq}
  \end{center}
\end{figure}

\begin{figure}[tb]
  \begin{center}
    \leavevmode
    \epsfxsize=11cm 
    \epsfbox{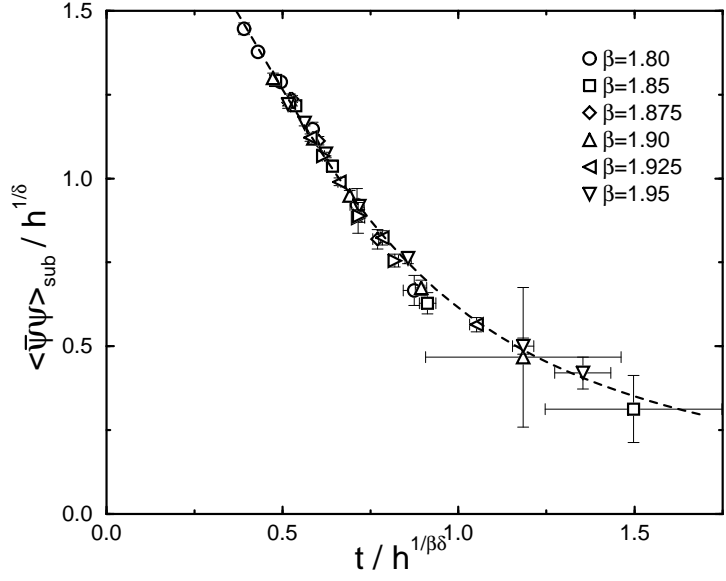}
    \caption{O(4) scaling fit with $h=2 m_q a$, 
    using all data at $\beta=1.8$--1.95
    and $2m_q a < 0.9$ on $16^3\times 4$ lattice. 
    The best fit is obtained at $\beta_{ct}=1.469$.}
    \label{fig:o4scal}
  \end{center}
\end{figure}

Figure~\ref{fig:pbp-2mq} shows the results of 
$\langle\bar{\psi}\psi\rangle_{\rm sub}$ as a function of Ward-Takahashi 
identity quark mass from our action at $\beta=1.8$--2.2 obtained 
on a $16^3\times 4$ lattice.
We perform a fit to the O(4) scaling function \cite{Toussaint} 
by adjusting the value of $\beta_{ct}$ as well as the scales for 
$t$ and $h$, with the exponents fixed to the O(4) values \cite{O4exponent}.
Since scaling is expected toward the chiral transition point corresponding to 
massless quark, we start our fit with the entire data set, gradually reducing 
the range of $(\beta, m_qa)$ toward smaller $\beta$ and $m_qa$ 
until an acceptable $\chi^2/N_{DF}$ is obtained. 
We find this condition to become satisfied for the range 
$\beta=1.8$--1.95 and $2m_q a=0.0$--0.9 
with $\chi^2/N_{DF}=0.82$ for $N_{DF}=33$,    
for which we obtain 
\be
\beta_{ct} = 1.469(73).  
\ee
This fit is shown in Fig.~\ref{fig:o4scal}. 

\begin{figure}[tb]
  \begin{center}
    \leavevmode
    \epsfxsize=11cm 
    \epsfbox{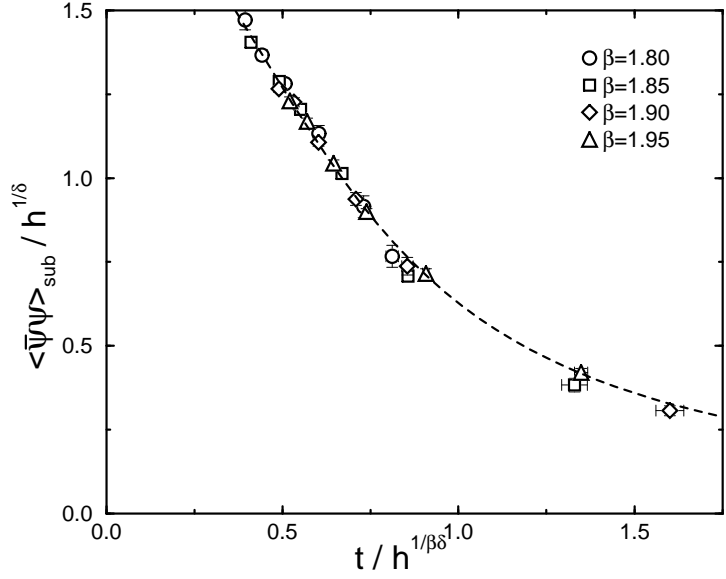} 
    \caption{The same as Fig.~\ref{fig:o4scal} using
    $m_q a$ obtained on the zero-temperature $16^4$ lattice.
    The best fit is obtained at $\beta_{ct}=1.462$.}
    \label{fig:o4scal0}
  \end{center}
\end{figure}

In this test we use results for $m_q a$ obtained on the finite-temperature
$16^3\times 4$ lattice.  
We repeat the test using $m_q a$ determined on the zero
temperature lattice, $16^4$, at the same values of $(\beta,K)$. 
Results for quark mass extracted on a $16^4$ and a $16^3\times 4$ lattice
are mutually consistent for the range $\beta=1.8$--1.95 and $2m_q a=0.0$--0.9. 
As shown in Fig.~\ref{fig:o4scal0},
we again find a good agreement with the O(4) scaling function 
with $\chi^2/N_{DF}=1.18$ for $N_{DF}=22$
at 
\be
\beta_{ct} = 1.462(66).
\ee
The number of data points are smaller due to fewer 
simulation points on the $16^4$ lattice. 

The good consistency of our condensate data with the O(4) scaling suggests 
that the chiral phase transition of two-flavor QCD in the continuum limit
is of second order. 
The estimated value for $\beta_{ct}$ is somewhat low compared to the result 
for the cusp $\beta_{\rm cusp}=\betacusp$ obtained in Sec.~\ref{sec:PBphase}. 
The reason for this trend is not clear at present. A possible origin is 
lack of condensate data below $\beta=1.8$, closer toward the chiral 
transition point, in the scaling fit. 
We leave an examination of this point for future work.

\section{Scaling of the pseudo-critical transition point}
\label{sec:scalingB}

Let us denote by $t_{pc}$ the pseudo-critical point 
defined as the peak position of the magnetic susceptibility 
at finite $h$. 
From the scaling relation (\ref{eq:o4}), we expect 
\be
t_{pc} \propto h^{1/\beta\delta} 
\label{eq:tpc}
\ee
to be satisfied for this quantity. 
Identifying $t_{pc} = \beta_{pc} - \beta_{ct}$,
where $\beta_{pc}$ is the $\beta$-coordinate of the $K_t$-line, 
we expect
\be
\beta_{pc} 
= \beta_{ct} + B_{\beta} h^{1/\beta\delta}\left( 1+O(h)\right)
\label{eq:BpcFit}
\ee
near the chiral transition point.

\begin{figure}[tb]
  \begin{center}
    \leavevmode
    \epsfxsize=11cm 
    \epsfbox{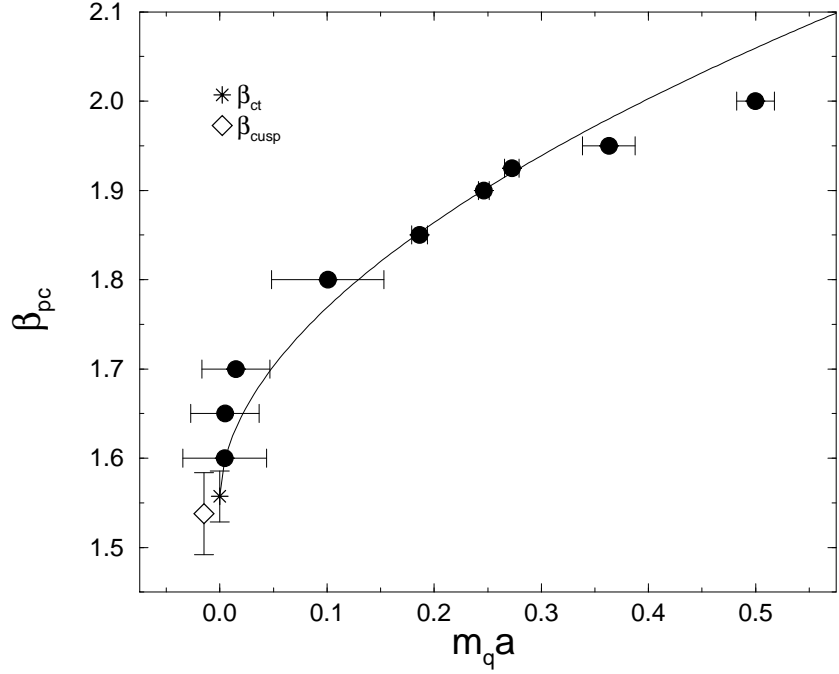}
    \caption{
	The pseudo-critical point $\beta_{pc}$ as a function of 
	the Ward identity quark mass $m_qa$.
	Result of the fit using the O(4) exponent is shown by a solid 
        line.
	The prediction for $\beta_{ct}$ in the chiral limit from
	this fit is shown by a star.
	The location of the cusp of the parity-broken phase is shown by a 
	diamond, which is slightly shifted in the horizontal direction for
	clarity of the figure.}
    \label{fig:Mqfit}
  \end{center}
\end{figure}

In Fig.~\ref{fig:Mqfit} we plot the pseudo critical point $\beta_{pc}$ 
choosing
\be
h=m_qa
\ee
{\it i.e., } the quark mass defined by the axial vector Ward-Takahashi 
identity.  
Errors attached to $m_qa$ include the uncertainty in the location 
of the $K_t$ line for each value of $\beta_{pc}$, which actually 
dominates the error.  
The precision of our results is not sufficient to attempt a fit 
taking the exponent $1/\beta\delta$ as a free parameter.  
We therefore make a fit adopting the $O(4)$ value for this exponent. 
Taking  $1.6\leq\beta_{pc}\leq 1.95$ for the fit range, since this was 
the range acceptable for the scaling analysis of the chiral condensate, 
we obtain 
\be
\beta_{ct}=1.557(28)
\label{eq:bctmq}
\ee
with $\chi^2/N_{DF}=1.04$ for $N_{DF}=6$.  

\begin{figure}[tb]
  \begin{center}
    \leavevmode
    \epsfxsize=11cm 
    \epsfbox{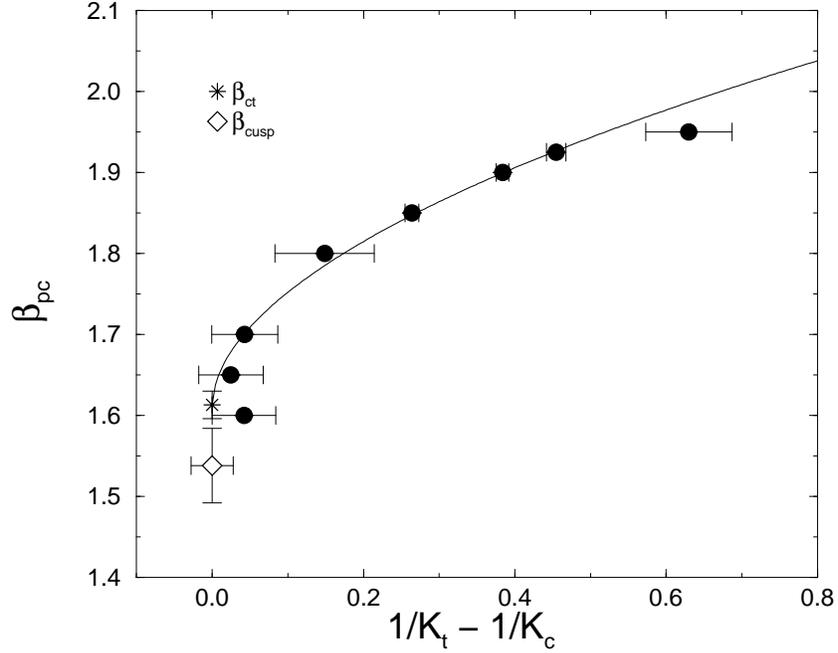}
    \caption{The same as Fig.~\ref{fig:Mqfit} for $h=1/K_t-1/K_c$.}
    \label{fig:Ktfit}
  \end{center}
\end{figure}

In Fig.~\ref{fig:Ktfit} we show results of a similar analysis, 
making an alternative choice,
\be
h = 1/K_t - 1/K_c,
\ee
for the external field.  The fit using this identification 
leads to 
\be
\beta_{ct}=1.613(17) 
\label{eq:bctK}
\ee
for data in the same range $1.6\leq\beta\leq 1.95$ with $\chi^2/N_{DF}=1.60$ 
for $N_{DF}=6$.  

\begin{figure}[tb]
  \begin{center}
    \leavevmode
    \epsfxsize=11cm 
    \epsfbox{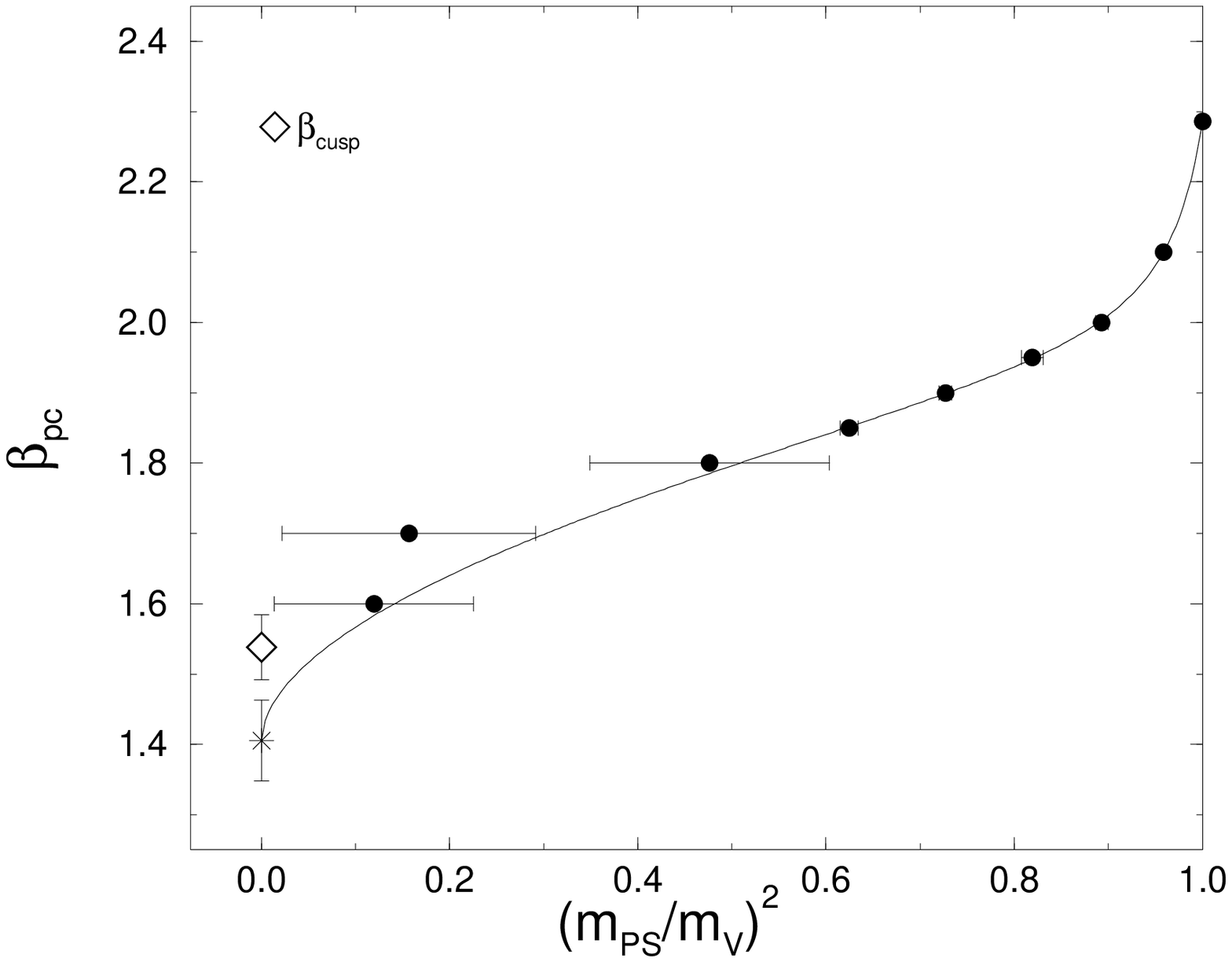}
    \caption{The same as Fig.~\ref{fig:Mqfit} for 
	$h=(m_{\rm PS}/m_{\rm V})^2$.
	The data $\beta_{pc}=\betaquench$ at $h=1$ is from 
	the pure gauge theory \protect\cite{kanekoTc,Okamoto}.}
    \label{fig:PSVfit}
  \end{center}
\end{figure}

As the third possibility, we consider
\be
h = (m_{\rm PS}/m_{\rm V})^2
\label{eq:hPSV}
\ee
obtained on the zero-temperature lattice.
This choice has the nice feature that the location 
of the physical point can be easily identified as 
$h = (m_{\pi}/m_{\rho})^2 \simeq 0.031$.
Furthermore the entire range of quark mass from $0$ to $\infty$ 
can be parameterized in a finite interval $0\leq h\leq 1$. 
Taking advantage of the latter feature, 
we attempt a global fit including
the data in the quenched limit \cite{kanekoTc,Okamoto}.
We find that a Pad\'e-type ansatz extending (\ref{eq:tpc}) given by 
\be
\beta_{pc} = c_\beta^{(0)} + c_\beta^{(1)} h^{1/\beta\delta} 
\frac{1 + c_\beta^{(2)} h}{1 + c_\beta^{(3)} h},
\label{eq:fitB}
\ee
reproduces our data well, as shown in Fig.~\ref{fig:PSVfit}, with 
$\chi^2/N_{DF}=0.33$ for $N_{DF}=5$.
The fitted value for the chiral transition point $\beta_{ct}=1.41(6)$ 
is too low, however, presumably because results toward larger values of $h$ 
including that in the quenched limit $(h=1)$, having 
small errors, dominantly determine the fit. 

\section{Chiral transition temperature}
\label{sec:scalingT}

\begin{figure}[tb]
  \begin{center}
    \leavevmode
    \epsfxsize=11cm 
    \epsfbox{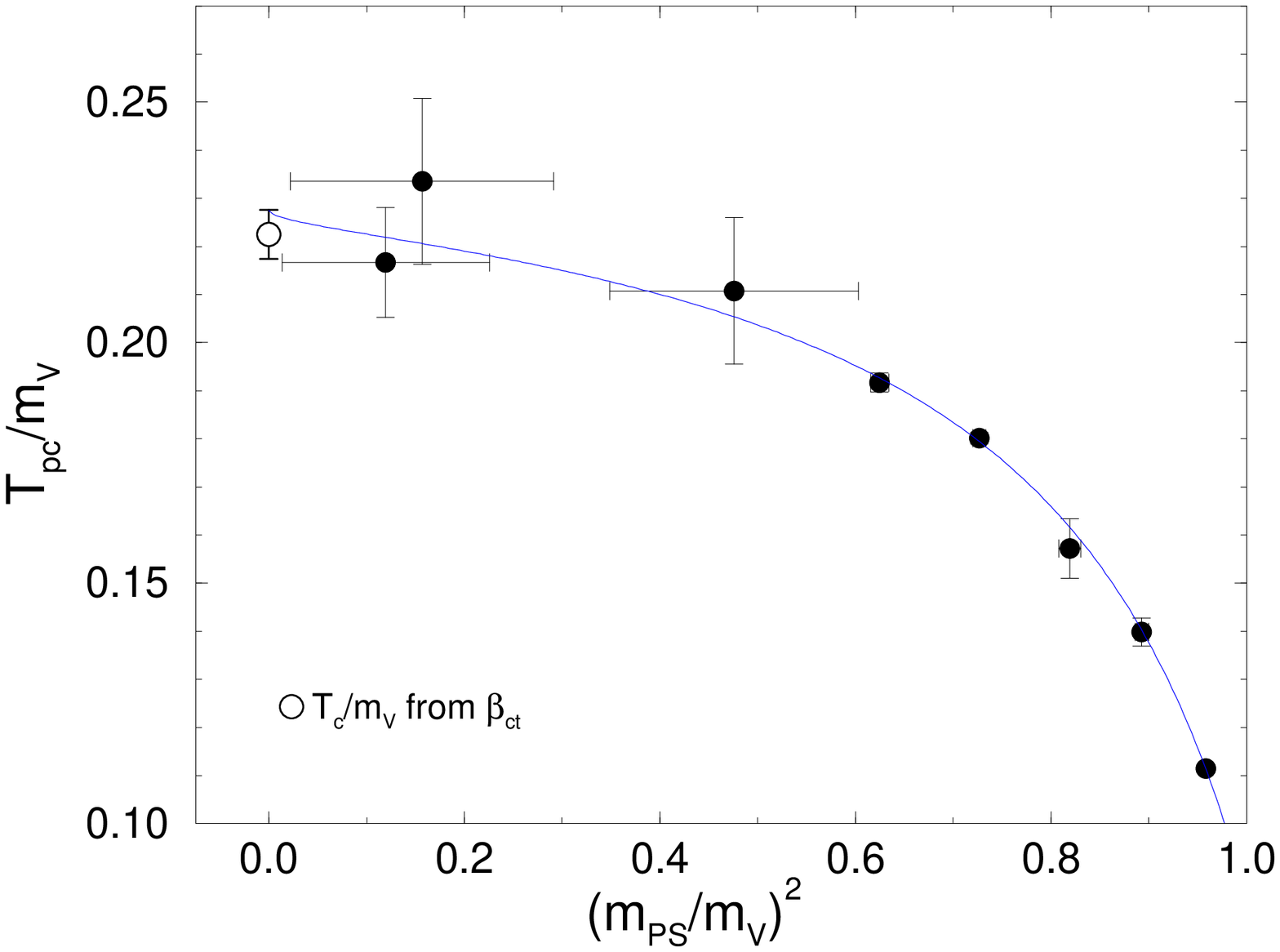}
    \caption{Chiral transition temperature in units of vector meson mass 
	as a function of $(m_{\rm PS}/m_{\rm V})^2$. 
	Hadron masses are measured at the same simulation point 
	on a zero-temperature lattice. 
	The chiral transition temperature estimated from $\beta_{ct}$
	discussed in Sec.~\ref{sec:scalingB} is shown with an open symbol.
	The solid line is a guide to the eyes based on a Pad\'e-type ansatz.}
    \label{fig:Tpcfit}
  \end{center}
\end{figure}

In order to calculate the transition temperature, we evaluate 
$T_{pc}/m_{\rm V}=1/(N_tm_{\rm V}a)$ where $m_{\rm V}$ is the 
vector meson mass at the transition point $(\beta, K_t(\beta))$ 
evaluated at zero temperature. 
The results are plotted in Fig.~\ref{fig:Tpcfit} 
as a function of $(m_{\rm PS}/m_{\rm V})^2$.   
We see that the difference in the values of $T_{pc}/m_{\rm V}$
at the chiral transition point $\beta_{ct}$ where 
$(m_{\rm PS}/m_{\rm V})^2 =0$ and at the physical point 
$(m_{\rm PS}/m_{\rm V})^2 =0.031$ is negligible compared with 
the current magnitude of errors.
Our values of $T_{pc}/m_{\rm V}$ are slightly smaller than those 
obtained with the standard Wilson quark action on $N_t=4$ lattices, 
and consistent with the results from other improved actions 
\cite{Karsch99}.

To estimate the transition temperature in the chiral limit,  
we interpolate the vector meson mass along the zero-temperature 
critical line $K_c(T=0)$ to $\beta=\beta_{ct}$. 
For the numerical value of $\beta_{ct}$ we take the result 
(\ref{eq:bctmq}) based on O(4) scaling of the $K_t$ line in terms 
of the Ward-Takahashi identity quark mass. 
This gives
\be
T_c/m_{\rm V} = 0.2224(51),
\label{eq:tcvfinal}
\ee
which is plotted by open circle in Fig.~\ref{fig:Tpcfit}.  
Converting to physical units, we obtain
\be
T_c = \tcfinalfinal \;{\rm MeV}
\label{eq:tcphysfinal}
\ee
The solid line drawn in Fig.~\ref{fig:Tpcfit} is a guide to the eyes, 
representing a global fit of the data to a Pad\'e type ansatz (\ref{eq:fitB}), 
with $\beta_{pc}$ replaced by $T_{pc}/m_{\rm V}$ and employing  
the O(4) values for the exponents.

\section{Conclusions}
\label{sec:conclusion}

We have studied the phase structure and the nature of the 
chiral transition in two-flavor QCD at finite temperatures 
using an RG-improved gauge action and a clover-improved 
Wilson-type quark action 
on a lattice with the temporal size $N_t=4$. 
We have identified the boundary of the parity-broken phase and 
the finite temperature transition line on a two-parameter space of
the coupling and the hopping parameter.
The chiral transition point is found to be located very close to 
the cusp point of the parity-broken phase. 

A subtracted chiral condensate is shown to satisfy the scaling 
behavior with the exponents and the scaling function universal to the 
O(4) Heisenberg model.
The quark mass dependence of the transition point is also 
consistent with the O(4) prediction.
These results, in agreement with the previous study of the 
standard Wilson quark action combined with the RG-improved gauge 
action \cite{Tsukuba97,Tsukuba97a}, 
indicate that the chiral transition of two-flavor QCD is of second order 
in the continuum limit.

Assuming the O(4) scaling, we have extrapolated the transition 
point towards the chiral limit.
Fixing the lattice scale in terms of the $\rho$ meson mass, 
we obtain $T_c = \tcfinalfinal$ MeV for the transition temperature.

\section*{Acknowledgments}

This work is supported in part by Grants-in-Aid of the Ministry of 
Education (Nos.~09304029, 10640246, 10640248,
10740107, 11640250, 11640294, 11740162, 12014202, 12640253). 
AAK and TM are supported by the Research for Future Program of JSPS
(No. JSPS-RFTF 97P01102). 
SE, KN, M. Okamoto, and HPS are JSPS Research Fellows. 


\begin{table} [htb]
\begin{center}
  \begin{tabular}{cccc}
$\beta$ & $K$ & traj. & therm. \\ \hline
1.800 & 0.1300--0.1450 &  500--2000 & 200--500 \\ 
1.825 & 0.1425 & 1000 & 300 \\ 
1.850 & 0.1250--0.1440 &  500--1900 & 200--300 \\ 
1.865 & 0.1400 & 2000 & 300 \\ 
1.875 & 0.1350--0.1400 & 1580--2000 & 200 \\ 
1.890 & 0.1400 & 2200 & 300 \\ 
1.900 & 0.1250--0.1425 &  500--2000 & 200--400 \\ 
1.910 & 0.1350 & 2100 & 300 \\ 
1.925 & 0.1300--0.1400 & 1000--2000 & 200--300 \\ 
1.950 & 0.1200--0.1410 &  500--2000 & 200 \\ 
1.975 & 0.1300 & 1000 & 200 \\ 
2.000 & 0.1150--0.1390 &  500--2000 & 200--300 \\ 
2.100 & 0.0900--0.1375 &  500--1000 & 200--900 \\ 
2.200 & 0.0700--0.1365 &  500 & 200 \\ 
\end{tabular}
\end{center}
\caption{Simulation parameters on the $16^{3} \times 4$ lattice.}
\label{tab:param16x4}
\end{table}

\begin{table} [htb]
\begin{center}
  \begin{tabular}{cccc}
$\beta$ & $K$ & traj. & therm. \\ \hline
1.80 & 0.1300--0.1450 & 200 & 200--500 \\ 
1.85 & 0.1250--0.1440 & 200--300 & 100--300 \\ 
1.90 & 0.1250--0.1425 & 200 & 200--400 \\ 
1.95 & 0.1200--0.1410 & 200--300 & 100--400 \\ 
2.00 & 0.1150--0.1390 & 200--300 & 100--200 \\ 
2.10 & 0.0900--0.1375 & 300 & 200--550 \\ 
2.20 & 0.0700--0.1365 & 200--300 & 100--200 \\ 
2.25 &    0.1300--0.1360 & 250 &      \\ 
\end{tabular}
\end{center}
\caption{Simulation parameters on the $16^{4}$ ($\beta=1.8$--2.2)
         and $16^{3} \times 42$ ($\beta = 2.25$) lattice.}
\label{tab:param16x16}
\end{table}

\begin{table} [htb]
\begin{center}
  \begin{tabular}{cccc}
$\beta$ & $K$ & traj. & therm. \\ \hline
 1.20  &   0.1425--0.1475 &  118--120  &  70--72  \\
 1.30  &   0.1475--0.1525 &  118--120  &  70--104 \\
 1.40  &   0.1500--0.1540 &  120--180  &  110--280 \\
 1.40  &   0.1625--0.1750 &  84--100   &  36--80   \\
 1.50  &   0.1475--0.1540 &  100--120  &  70      \\
 1.50  &   0.1585--0.1700 &  88--100   &  50--90  \\
 1.60  &   0.1400--0.1525 &  250--300  &  75--510 \\
 1.60  &   0.1560--0.1700 &  92--500   &  50--300 \\
 1.65  &   0.1400--0.1500 &  300--500  &  200--460 \\
 1.65  &   0.1565--0.1700 &  250--500  &  100--400  \\
 1.70  &   0.1400--0.1500 &  250--300  &  150--300  \\
 1.70  &   0.1545--0.1700 &  250       &  80--240   \\
\end{tabular}
\end{center}
\caption{Simulation parameters on the $8^{3} \times 4$ lattice.}
\label{tab:param8x4}
\end{table}

\begin{table} [htb]
\begin{center}
  \begin{tabular}{cccc} 
$\beta$ & $K$ & traj. & therm. \\ \hline
   1.30 &    0.1375--0.1500 & 120  & 55--70  \\ 
   1.40 &    0.1400--0.1525 & 90--120  & 50--70 \\ 
   1.50 &    0.1375--0.1525 & 90--120  & 50--80 \\ 
   1.60 &    0.1370--0.1510 & 700--1240 & 50--150  \\ 
   1.70 &    0.1340--0.1510 & 120--1275 & 150  \\ 
\end{tabular}
\end{center}
\caption{Simulation parameters on $12^{3} \times 24$ lattice.}
\label{tab:param12x24}
\end{table}

\begin{table}[htb]
\begin{center}
  \begin{tabular}{lllll}
 $\beta$ & $a^{-1}$ [GeV] 
& $K_c(T=0)$ & $K_c^{\rm lower}(N_t=4)$ & $K_c^{\rm upper}(N_t=4)$ \\ 
\hline
 1.20      &    &	& 0.15044(20) &              \\
 1.238(12) &	&	&             &  0.17000     \\
 1.30      & 0.563(21) & 0.154082(56) & 0.15475(25) &              \\
 1.303(07) &	&	&             &  0.16500     \\
 1.364(04) &	&	&             &  0.16250     \\ 
 1.40      & 0.606(31) & 0.156308(62) & 0.15702(15) &  0.16103(26) \\
 1.432(05) &	&	&             &  0.16000     \\
 1.50    & 0.658(11) & 0.156641(58) & 0.15730(22) &  0.15780(21) \\
 1.60    & 0.707(10) & 0.155259(31) &             &            \\  
 1.70    & 0.780(10) & 0.151987(22) &             &            \\  
 1.80(*) & 0.995(19) & 0.147678(15) &             &            \\  
 1.85    & 1.033(29) & 0.145526(58) &             &            \\  
 1.90    & 1.207(36) & 0.143737(48) &             &            \\  
 1.95(*) & 1.331(23) & 0.142072(14) &             &            \\  
 2.00    & 1.445(33) & 0.140811(55) &             &            \\  
 2.10(*) & 1.851(58) & 0.139020(21) &             &            \\  
 2.20(*) & 2.42(14)  & 0.137658(53) &             &            \\  
 2.25    & 2.40(24)  & 0.137225(92) &             &            \\
\end{tabular}
\end{center}
\caption{Results for the lattice scale $a^{-1}$, 
the chiral limit $K_c(T=0)$, 
and the upper and lower boundaries of the parity-broken phase $K_c(N_t=4)$.
Results for $K_c(T=0)$ and $a^{-1}$ at $\beta$ marked by (*) 
are obtained using hadron mass results from \protect\cite{CPPACSfull}.
}
\label{tab:Kc}
\end{table}

\begin{table}[hbt]
\begin{center}
  \begin{tabular}{llll}
 $\beta$ & $K_t(N_t=4)$ & $m_{\rm PS}/m_{\rm V}$ &  
$T_{pc}/m_{\rm V}$ \\ \hline
1.600 & 0.1543(10)  &  0.346(153) &  0.217(11) \\
1.650 & 0.1533(10)  &             &            \\
1.700 & 0.1510(10)  &  0.396(170) &  0.234(17) \\
1.800 & 0.1445(14)  &  0.690(92)  &  0.211(15) \\
1.850 & 0.14019(18) &  0.7905(60) &  0.1917(20) \\
1.900 & 0.13621(15) &  0.8525(39) &  0.1801(12) \\
1.925 & 0.13417(23) &             &             \\
1.950 & 0.13040(97) &  0.9051(64) &  0.1572(62) \\
2.000 & 0.12371(73) &  0.9450(36) &  0.1398(29) \\
2.100 & 0.10921(43) &  0.9790(13) &  0.1114(09) \\
\end{tabular}
\end{center}
\caption{Finite temperature transition/crossover point $K_t$
for $N_t=4$. 
Results for $m_{\rm PS}(T=0)/m_{\rm V}(T=0)$ 
and $T_{pc}/m_{\rm V}(T=0)$ interpolated to the $K_t$ point 
are also listed.
No zero-temperature simulations were made at $\beta=1.65$ and 1.925.
}
\label{tab:Kt}
\end{table}

\begin{thebibliography}{99}


\bibitem{Karsch99} 
For a recent review, see, 
F.\ Karsch, Nucl.\ Phys.\ B (Proc.\ Suppl.) 83-84 (2000) 14.

\bibitem{MILC-EOS} T. Blum {\it et al.}, Phys. Rev. D51 (1995) 5153;
C. Bernard {\it et al.}, Phys. Rev. D55 (1997) 6861.

\bibitem{Ukawa96} 
For a review and references, see, 
A. Ukawa, Nucl.\ Phys.\ B (Proc.\ Suppl.) 53 (1997) 106.

\bibitem{Okamoto} CP-PACS Collaboration, M. Okamoto {\it et al.}, 
Phys. Rev. D60 (1999) 094510.

\bibitem{Iwasaki83}
Y.\ Iwasaki, Nucl.\ Phys.\ B258 (1985) 141; 
Univ.\ of Tsukuba report UTHEP-118 (1983), unpublished.

\bibitem{Bielefeld}G. Boyd {\it et al.}, Nucl. Phys. B469 (1996) 419.

\bibitem{prelim}
A preliminary account of our results has been presented in, 
CP-PACS Collaboration, A.\ Ali Khan {\it et al.},
Nucl.\ Phys.\ B (Proc.\ Suppl.) 83-84 (2000) 360.

\bibitem{Bielefeld-EOS} 
J. Engels {\it et al.}, Phys. Lett. B396 (1997) 210;
U. Heller {\it et al.}, Phys. Rev. D60 (1999) 114502;
F. Karsch {\it et al.}, Phys. Lett. B478 (2000) 447.

\bibitem{MILC} 
C.\ Bernard {\it et al.}, Phys.\ Rev.\ D49 (1994) 3574; 
T.\ Blum {\it et al.}, Phys.\ Rev.\  D50 (1994) 3377.

\bibitem{Tsukuba96} 
Y.\ Iwasaki, K.\ Kanaya, S.\ Kaya, S.\ Sakai, and T.\ Yoshi\'e,
Phys.\ Rev.\  D54 (1996) 7010.

\bibitem{Tsukuba97}
Y.\ Iwasaki, K.\ Kanaya, S.\ Kaya, and T.\ Yoshi\'e,
Phys.\ Rev.\ Lett.\ 78 (1997) 179.

\bibitem{SWclover}
B.\ Sheikholeslami and R.\ Wohlert, Nucl.\ Phys.\ B259 (1985) 572.

\bibitem{cloverFT}
S.\ Aoki, A.\ Ukawa, T.\ Umemura, unpublished (1995),
refered in A.\ Ukawa, Nucl.\ Phys.\ B (Proc.\ Suppl.) 53 (1997) 106;
R.G.\ Edwards and U.M.\ Heller, Phys.\ Lett.\ B462 (1999) 132.

\bibitem{MILC97}
C.\ Bernard {\it et al.}, Phys.\ Rev.\ D56 (1997) 5584.

\bibitem{Aoki}
S.\ Aoki, Phys.\ Rev.\  D30 (1984) 2653; 
S.\ Aoki, Nucl.\ Phys.\ B (Proc.\ Suppl.) 60A (1998) 206.

\bibitem{AUU96}
S. Aoki, A. Ukawa, T. Umemura, Phys.\ Rev.\ Lett.\ 76 (1996) 873;
S. Aoki, T. Kaneda, A. Ukawa, T. Umemura,
Nucl. Phys. B(Proc. Suppl) 53 (1997) 438.

\bibitem{Tsukuba97a}
S.\ Aoki, Y.\ Iwasaki, K.\ Kanaya, S.\ Kaya, A.\ Ukawa, and T.\ Yoshi\'e,
Nucl.\ Phys.\ B (Proc.\ Suppl.) 63 (1998) 397.

\bibitem{PisarskiWilczek}
R.\ Pisarski and F.\ Wilczek, Phys.\ Rev.\ D29 (1984) 338;
F.\ Wilczek, Int.\ J.\ Mod.\ Phys.\ A7 (1992) 3911;
K.\ Rajagopal and F.\ Wilczek, Nucl.\ Phys.\ B399 (1993) 395.

\bibitem{KL} F. Karsch, Phys. Rev. D49 (1994) 3791; 
F. Karsch and E. Laermann, Phys. Rev. D50 (1994) 6954;
A.\ Berera, Phys.\ Rev.\ D50 (1994) 6494.

\bibitem{O4JLQCD}
S.\ Aoki {\it et al.}, Phys.\ Rev.\ D57 (1998) 3910.

\bibitem{O4Bielefeld}
E.\ Laermann, Nucl.\ Phys.\ B (Proc.\ Suppl.) 63 (1998) 114.

\bibitem{O4MILC}
C.\ Bernard {\it et al.}, Phys.\ Rev.\ D61 (2000) 054503.

\bibitem{comparative}
CP-PACS Collaboration, S.\ Aoki {\it et al.}, 
Phys.\ Rev.\ D60 (1999) 114508.

\bibitem{CPPACSfull}
R.\ Burkhalter for the CP-PACS Collaboration,
Nucl.\ Phys.\ B (Proc.\ Suppl.) 73 (1999) 3;
CP-PACS Collaboration, S.\ Aoki {\it et al.}, 
Nucl.\ Phys.\ B (Proc.\ Suppl.) 73 (1999) 192, 216;
A.\ Ali Khan {\it et al.}, Nucl.\ Phys.\ B (Proc.\ Suppl.) 83-84 (2000) 176.

\bibitem{CPPACSmq}
CP-PACS Collaboration, A.\ Ali Khan {\it et al.},
hep-lat/0004010, to be published in Phys. Rev. Lett.

\bibitem{csw-one-loop}
S.\ Aoki {\it et al.}, Nucl.\ Phys.\ B540 (1999) 501.

\bibitem{Itoh} 
S.\ Itoh, Y.\ Iwasaki, Y.\ Oyanagi and T.\ Yoshi\'e,
Nucl.\ Phys.\ B274 (1986) 33.

\bibitem{Bochicchio} 
M.\ Bochicchio, L.\ Maiani, G.\ Martinelli,
G.\ Rossi and M.\ Testa, Nucl.\ Phys.\ B262 (1985) 331. 

\bibitem{CPPACSq}
CP-PACS Collaboration, S.\ Aoki {\it et al.},
Phys.\ Rev.\ Lett.\ 84 (2000) 238.

\bibitem{kanekoTc}
Y.\ Iwasaki, K.\ Kanaya, T.\ Kaneko and T.\ Yoshi\'e, 
Phys.\ Rev.\  D56 (1997) 151.

\bibitem{O4exponent}
K.\ Kanaya and S.\ Kaya, Phys.\ Rev.\ D51 (1995) 2404;
P.\ Buetera and M.\ Comi, Phys.\ Rev.\ B52 (1995) 6185;
H.G.\ Ballesteros, L.A.\ Fern\'andez, V.\ Mart\'in-Mayor
and A.\ Mu\~noz Sudupe, Phys.\ Lett.\ B387 (1996) 125.

\bibitem{Toussaint}
D.\ Toussaint, Phys.\ Rev.\  D55 (1997) 362.
J.\ Engels and T.\ Mendes, hep-lat/9911028.

\end{thebibliography}
\end{document}